# Room-temperature polariton supersolids in organic-inorganic halide perovskites

Yuanhao Gong[1,2,*], Jingwen Ma[1,3,*], Shuang Zhang[1,3], Xiaobo Yin[1,2,†], Xiang Zhang[1,2,3,†]

[1]State Key Laboratory of Optical Quantum Material and Department of Physics, The University of Hong Kong, Hong Kong, China

[2]Department of Mechanical Engineering, The University of Hong Kong, Hong Kong, China

[3]Department of Electrical and Electronic Engineering, The University of Hong Kong, Hong Kong, China

[*]These authors contribute equally

[†]Correspondence to: (Y. X.) xbyin@hku.hk; (X. Z.) president@hku.hk

**Abstract:** Exploring exotic quantum phases of matter at room temperature represents a frontier challenge in modern physics. The supersolid phase, uniquely merging crystalline order with frictionless superfluid flow, stands among the most intriguing macroscopic quantum phenomena. To date, many demonstrations of supersolidity, whether in ultracold atomic gases or III–V semiconductor–based polariton systems, have been confined to cryogenic temperatures. Here, we report the observation of room-temperature supersolids in photonic-crystal polariton condensates. By integrating a room-temperature-stable organic-inorganic halide perovskite semiconductor with a dispersion-engineered photonic-crystal waveguide, we create a polariton condensate with multi-mode dispersion landscapes and pronounced parametric nonlinearities. Above a critical condensation density, the interacting condensates spontaneously break continuous translational symmetry, creating a non-rigid supersolid phase that simultaneously exhibits emergent crystalline order and global quantum coherence. This work establishes a room-temperature platform for investigating quantum hydrodynamics and developing coherent quantum simulation devices.

Macroscopic quantum coherent states[1], wherein a multitude of particles collectively behave as a single quantum wave, is a cornerstone of modern physics. It underpins many intriguing phenomena such as Bose-Einstein condensation (BEC)[2], superfluidity[3], and superconductivity[4]. The hallmark of these states is the off-diagonal long-range order (ODLRO) that embodies global quantum phase coherence[5]. This contrasts fundamentally with the diagonal long-range order (DLRO) that defines classical crystalline solids. It was this dichotomy that inspired the profound question posed by Penrose and Onsager in 1956[6]: can a single quantum phase simultaneously host the DLRO of a crystalline solid and the ODLRO of a condensate? This seminal inquiry marked the conceptual birth of the supersolid phase.

After decades of research in solid $^4He$[5,7-9], the first definitive experimental signature of supersolidity emerged in 2017 through experiments with ultracold atomic gases, which employed synthetic spin–orbit coupling[10] and cavity-mediated long-range interactions[11]. These demonstrations were soon extended to dipolar quantum gas systems[12-14], where intrinsic dipole–dipole interactions spontaneously break translational symmetry, organizing the condensate into a periodic droplet array while preserving global phase coherence. Subsequent work in these highly tunable platforms has further elucidated the fascinating properties of the supersolid phase through the observation of Goldstone modes[15,16], two-dimensional droplet arrays[17], and quantized vortices[18]. Nevertheless, all these demonstrations remain confined to the nanokelvin temperature regime, requiring complex cooling and vacuum apparatus.

Exciton–polaritons provide a promising route to overcome this temperature barrier, as their driven-dissipative nature can sustain transient macroscopic quantum coherence at elevated temperatures. Initially realized in III–V semiconductor quantum wells[19], these hybrid light-matter quasiparticles have emerged as a versatile platform for realizing polariton BECs[20-21] and superfluids[22]. The recent advent of novel materials with strong excitonic resonances, particularly organic semiconductors and halide perovskites, has further enabled room-temperature BECs and superfluids[23-35], demonstrating the capability to sustain ODLRO under ambient conditions. Very recently, the supersolid phase has been realized in photonic-crystal exciton–polariton condensate[36-37] using III–V semiconductor quantum-wells, where the tailored polariton dispersion provides predefined multi-mode channels for continuous translational symmetry breaking. Realizing a supersolid at room temperature, which would establish a practical, on-chip platform for exploring quantum many-body physics and hydrodynamics under ambient conditions, then becomes a booming research direction[38-40]. To date, direct observation of polariton supersolids in organic or organic-inorganic hybrid systems has remained elusive, this is partly due to the inherent instability of the material itself, as well as the limitations of the growth and transfer methods[41-42].

In this work, we report the experimental realization of a room-temperature polariton supersolid in an organic-inorganic hybrid photonic-crystal device. We construct the device by integrating a thin-film perovskite semiconductor methylammonium lead bromide ($MAPbBr_3$) into a dispersion-engineered multi-mode photonic-crystal waveguide. The large exciton binding energy in $MAPbBr_3$, unlike III–V semiconductors, enables the formation of stable polariton condensates at ambient conditions. Above a critical condensate density, we observe the spontaneous emergence of a room-temperature supersolid, driven by parametric scattering within the engineered multi-mode polariton dispersions. This supersolid is intrinsically non-rigid, evidenced by a continuous shift of its crystalline wavevector with increasing condensate density. We provide definitive experimental evidence of supersolidity by directly visualizing the emergent DLRO through real-space microscopy and verifying the ODLRO via momentum-space quantum coincidence measurements. This work establishes room-temperature exciton–polariton photonic crystals as a versatile and scalable testbed for quantum many-body physics and opens avenues for practical coherent quantum devices operating under ambient conditions.

We experimentally demonstrate the room-temperature supersolid phase using a tailored exciton–polariton device, schematically shown in Fig. 1a. The platform is constructed by integrating a thin film of $MAPbBr_3$ into a dispersion-engineered photonic-crystal waveguide. The $MAPbBr_3$ films are grown using a space-confined method (Supplementary Information Sec. 1.1 and Fig. S1), with lateral dimensions of several tens of micrometers and typical thicknesses of a few hundred nanometers. These films exhibit a sharp excitonic resonance at $\hbar\omega_e = 2.377$ eV (Fig. S5), enabling robust polariton condensation at room temperature (Supplementary Information Sec. 4.1). Compared to common inorganic perovskite semiconductors, such as $CsPbCl_3$ and $CsPbBr_3$, organic cations (such as $MA^+$) in hybrid perovskites exhibit dynamic disorder that passivates defects and reduces trap depth, thereby suppressing nonradiative recombination[43-46]. By performing angle-resolved optical microscopy and analyzing the transmission spectrum, we identify and select a film with uniform thickness of $t_{\mathrm{pvk}} = 70$ nm for integration into the photonic crystal. Here, we emphasize the advantages of our integration method, the mechanical fragility of perovskite crystals renders their transfer onto other substrates susceptible to fracture. In addition, the heating and immersion in organic polar solvents during the micro-transfer method can damage the structure of perovskite crystals to some extent. Our method preserves the exciton properties of organic-inorganic hybrid perovskites to the greatest extent possible.

The dispersion-engineered photonic crystal is patterned into a silicon nitride ($Si_3N_4$) thin film (Fig. 1a, inset), with key device parameters as follows: $Si_3N_4$ etch depth $t_{etch}$ = 50 nm, residual slab substrate thickness $t_{sub}$ = 200 nm, lattice constant $a$ = 290 nm, and filling factor $r$ = 0.2. This hybrid structure supports strongly coupled exciton–polariton waveguide modes, including the fundamental $TE_0$ and first-order $TE_1$ branches. The photonic crystals fold the fundamental lower-polariton branch to the Brillouin zone center ($k_x = k_y = 0$), creating a high-quality-factor ($Q$), symmetry-protected bound state in the continuum (BIC), which we label as $\psi_0$ (Fig. 1b). Concurrently, it engineers the first-order lower-polariton branch to form a symmetric pair of counter-propagating modes, $\psi_1(\pm\mathbf{k})$, at finite momenta about the Brillouin zone center. A slight structural asymmetry along the vertical ($z$) direction induces weak hybridization between the fundamental and first-order waveguide modes. Nevertheless, in the ideal limit of a perfectly periodic structure, the Bloch modes $\psi_0$ and $\psi_1(\pm\mathbf{k})$ remain orthogonal in momentum space due to their distinct wavevectors. This momentum-space orthogonality allows the modes to be populated independently and facilitates non-degenerate nonlinear interactions between them.

The supersolid phase emerges through a two-threshold process, which we trace in momentum space using an angle-resolved optical microscope (Supplementary Information Sec. 3.4, Fig. S4). We pump the device using a femtosecond laser at central wavelength of 500 nm. At a low pump fluence of 10.38 $\mu J/cm^2$, the device exhibits broad spontaneous emission from multiple polariton branches (Fig. 2a). The band-edge BIC mode possesses a negative effective mass, which facilitates polariton accumulation with increasing pump intensity. As the pump fluence reaches the first threshold, $P_{th}^{BEC}$ = 13.14 $\mu J/cm^2$, a sharp condensate forms in the high-$Q$ BIC mode $\psi_0$ (Fig. 2b). Its emission at $k_x = 0$ is suppressed due to the BIC mode's far-field symmetry. Crucially, when the pump fluence exceeds a second, higher threshold at $P_{th}^{SS}$ = 15.89 $\mu J/cm^2$, distinct spectral peaks emerge at finite momenta, signifying macroscopic occupation of the first-order modes $\psi_1(\pm\mathbf{k}_{ss})$ (Fig. 2c). This macroscopic occupation of energy-degenerate, counter-propagating modes signals the breaking of continuous translational symmetry and marks the onset of the supersolid phase.

Figure 3a quantifies this threshold behavior by plotting the integrated intensities of the $\psi_0$ and $\psi_1(\pm\mathbf{k}_{ss})$ modes as a function of pump intensity. Condensation into the efficiently populated $\psi_0$ mode occurs at the first threshold $P_{th}^{BEC}$, followed by the emergence of a macroscopic population in the $\psi_1(\pm\mathbf{k}_{ss})$ modes at the second threshold, $P_{th}^{SS}$. Figure 3b shows the growth of the supersolid component density $n_1$ as a function of the condensate density $n_0$. Below the condensation threshold, in the spontaneous emission regime, $n_1$ scales as $n_1 \propto n_0^{0.85}$, indicating that the band-edge $\psi_0$ state is populated more efficiently than the $\psi_1(\pm\mathbf{k}_{ss})$ states. Between the two thresholds, $n_1$ exhibits sub-linear growth, scaling as $n_1 \propto n_0^{0.48}$, confirming that the population of $n_1$ originates from a nonlinear scattering process from the condensate. Above $P_{th}^{SS}$, the scaling becomes nearly linear, with $n_1 \propto n_0^{0.97}$. These power-dependent observations provide direct evidence that the supersolid is formed via nonlinear parametric scattering from condensate, rather than by linear processes.

We conduct theoretical analysis on the emergence of this supersolid (Supplementary Information Sec. 3). When sufficiently populated, the incoherent exciton reservoir drives a polariton condensate in the high-$Q$ $\psi_0$ mode. We treat this condensate as a classical coherent field, $\psi_0 = \sqrt{n_0} \cdot \exp(i\varphi_0)$, where $n_0$ and $\varphi_0$ suggest condensate density and phase, respectively. This macroscopic condensate mediates a squeezing interaction between the $\psi_1(\pm\mathbf{k})$ modes. The effective Hamiltonian governing the nonlinear dynamics of the $\psi_1(\pm\mathbf{k})$ modes is:

$$H = H_{\text{linear}} + \alpha_1 n_0 e^{2i\varphi_0} \psi_1^\dagger(\mathbf{k}) \psi_1^\dagger(-\mathbf{k}) + h.c.. \quad (1)$$

Here, $H_{\text{linear}} = [\Delta(\mathbf{k})\psi_1^\dagger(\mathbf{k})\psi_1(\mathbf{k}) + \Delta(-\mathbf{k})\psi_1^\dagger(-\mathbf{k})\psi_1(-\mathbf{k})]/2$ represents the non-interacting single-particle Hamiltonian, while the second term describes the parametric squeezing of the counter-propagating $\psi_1(\pm\mathbf{k})$ modes. In Eq. 错误!未找到引用源。, $\Delta(\mathbf{k}) = \Delta(-\mathbf{k})$ is the energy detuning of $\psi_1(\pm\mathbf{k})$ relative to the $\psi_0$ condensate, and $\alpha_1$ is the inter-modal nonlinear interaction coefficient between $\psi_0$ and $\psi_1(\pm\mathbf{k})$, which is determined by the Hopfield coefficients of the involved modes. The squeezing interaction in Eq. 错误!未找到引用源。 provides parametric gain to the $\psi_1(\pm\mathbf{k})$ modes. This gain is maximized at zero detuning ($\Delta(\mathbf{k}_{ss}) = \Delta(-\mathbf{k}_{ss}) = 0$), where its magnitude is $\alpha_1 n_0$. Here, $\mathbf{k}_{ss}$ denotes the specific wavevector satisfying the zero-detuning condition $\Delta(\mathbf{k}_{ss}) = \Delta(-\mathbf{k}_{ss}) = 0$. This theoretical analysis agrees well with the experimental observation of the zero-detuning behavior in Fig. 2c. When this gain exceeds the intrinsic dissipation rate $\Gamma(\mathbf{k}_{ss})$ of the $\psi_1(\pm\mathbf{k}_{ss})$ modes, i.e. $n_0 > \Gamma(\mathbf{k}_{ss})/\alpha_1$, a supersolid phase emerges spontaneously. This supersolid phase is characterized by macroscopic occupation of the first-order modes, with a density $n_1 = |\psi_1(\pm\mathbf{k}_{ss})|^2 \propto \sqrt{\alpha_1^2 n_0^2 - \Gamma^2(\mathbf{k}_{ss})}$. Notably, when $\alpha_1 n_0 \gg \Gamma(\mathbf{k}_{ss})$, the scaling approaches a linear dependence $n_1 \propto \alpha_1 n_0$, which is consistent with the experimental results in Fig. 3b.

Figure 3c presents the energy spectrum measured at the finite wavevector $\pm\mathbf{k}_{ss}$ for different condensation density $n_0$. A macroscopic population in the $\psi_1(\pm\mathbf{k}_{ss})$ states emerges once the pump intensity exceeds $P_{th}^{SS}$. With increasing $n_0$, the repulse polariton–polariton interactions induce a measurable blueshift of the emission energy. Because the $\psi_1(\pm\mathbf{k})$ modes exhibit a slanted dispersion, this energy shift directly corresponds to a change in the resonant wavevector $\mathbf{k}_{ss}$. As shown in Fig. 3d, the momentum-space distribution of the $\psi_1(\mathbf{k})$ state gradually shifts toward smaller $\mathbf{k}_{ss}$ as $n_0$ rises. This continuous tunability of the spatial period with condensate density demonstrates the non-rigid, interaction-driven nature of the supersolid phase.

Real-space imaging offers direct evidence for the DLRO characteristic of supersolidity. Above the supersolid threshold $P_{th}^{SS}$, the coherent superposition of the macroscopically occupied $\psi_0$ and $\psi_1(\pm\mathbf{k}_{ss})$ modes produces a periodic density modulation in the polariton emission (Fig. 4a). The period of this modulation is measured to be 3.33 μm, which aligns with the estimated value of $2\pi/|\mathbf{k}_{ss}| = 3.46$ μm. Notably, this density modulation is distinct from and incommensurate with the underlying photonic crystal lattice constant ($a$ = 290 nm), proving that crystalline order is not a trivial imprint of the patterned structure but arises spontaneously from nonlinear interactions. Supersolid formation spontaneously breaks a U(1) symmetry. In an ideal, infinite system, this results in a random relative phase between the $\psi_1(\mathbf{k}_{ss})$ and $\psi_1(-\mathbf{k}_{ss})$ modes for each realization of condensate. Resolving this random phase would therefore require single-shot measurements. In practice, the finite spatial extent of the pump spot, the photonic crystal, and the condensate itself introduces a confining potential, this finite-size effect induces a weak linear coupling between $\psi_1(\mathbf{k}_{ss})$ and $\psi_1(-\mathbf{k}_{ss})$, thereby locking their relative phase (Supplementary Information Sec. 3.4). Other mechanisms, such as linear scattering from $\psi_0$ condensate to the $\psi_1(\pm\mathbf{k}_{ss})$ modes, can also contribute to phase locking[26]. These effects enable the observation of a stable, time-averaged phase in our experiments, which integrate over million realizations. We note that analogous finite-size effects are inherent to atomic supersolids, yet they have not precluded the observation of key supersolid properties, such as well-defined Goldstone modes.

To characterize the ODLRO of the supersolid phase, we measure the second-order temporal coherence in momentum space using a Hanbury-Brown-Twiss (HBT) interferometer. Specifically, we use two fiber-coupled single-photon-counting detectors to measure the cross-correlation $g^2(\mathbf{k}_{ss}, -\mathbf{k}_{ss}; \tau)$ between the emission at $\pm\mathbf{k}_{ss}$. The fiber cores are precisely aligned to collect light exclusively from these two momentum components. Figure 4b shows the measured cross-correlation recorded far above the supersolid threshold. At zero time delay, the cross-correlation $g^2(\mathbf{k}_{ss}, -\mathbf{k}_{ss}; 0) = 1.0121$ is very close to 1, providing clear evidence for quantum coherence between $\psi_1(\mathbf{k}_{ss})$ and $\psi_1(-\mathbf{k}_{ss})$ modes. Combined with direct real-space observation of DLRO (Fig. 4a), these results provide definitive proof for the realization of supersolid phase at room temperature.

In conclusion, we experimentally realize a room-temperature supersolid in a dispersion-engineered photonic-crystal polariton device in organic-inorganic halide perovskites. Above a critical condensation density, the interacting condensates spontaneously break the continuous translational symmetry via isoenergetic parametric scattering, creating a robust supersolid phase at room temperature. We further visualize the DLRO through real-space microscopy and verify ODLRO via momentum-space quantum coincidence measurements.

This work establishes a practical platform for exploring the quantum hydrodynamics of supersolidity at room temperatures. The design flexibility of photonic crystals opens routes to engineer rich crystalline geometries, such as two-dimensional hexagonal or square supersolid, and to study associated topological defects like supersolid vortices. Furthermore, the non-rigid character of the supersolid implies the possibilities to observe Goldstone modes in its low-energy excitation spectrum[47]. Our device is uniquely suited for room-temperature spectroscopy of these collective modes via recently developed coherent probe spectroscopy[48], which will provide direct access to the superfluid stiffness and elastic properties of supersolids, enabling a complete hydrodynamic characterization. From a quantum photonics perspective, the intrinsic parametric scattering that drives supersolid formation constitutes an efficient on-chip non-classical light source. Unlike conventional squeezed light relying on an external, coherent parametric pump, our device utilizes a spontaneously generated polariton condensate[49]. Combined with the spectral tunability of organic-inorganic perovskite semiconductors and the versatile integration scheme, this platform opens avenues to generate squeezed and entangled light across various wavelengths, creating new opportunities for on-chip quantum photonic technologies.

## ACKNOWLEDGEMENTS

This work was supported by the Research Grants Council of Hong Kong under Grant N_HKU750/22 (X.Z.) and General Research Grant 17208725 (J.M.). The authors acknowledge the support in device nanofabrication from Prof. Yahui Xue and the Core Research Facilities at Southern University of Science and Technology.

## AUTHOR CONTRIBUTION STATEMENT

Y.G. and J.M. contribute equally to theoretical modeling, device fabrication, optical characterization, data analysis, and manuscript writing, with input from all other authors. X.Y. and X.Z. supervised the project.

## COMPETING INTERESTS STATEMENT

Authors declare no competing interests.

## METHODS

**Material synthesize.** The $MAPbBr_3$ single crystal films are synthesized using an optimized space-confined method. The process begins with preparing precursor solution by dissolving high-purity (99.99%) MABr and $PbBr_2$ powders in dimethyl formamide solvent at a 1:1 molar ratio. The solution is adjusted to 0.5 mol/L. In parallel, two ultrasonically cleaned quartz substrates are clamped together under external stress to form a tightly confined assembly. The precursor solution is then dispensed onto the edge of this assembly and drawn into the narrow gap by capillary action. Crystallization proceeds at 40 °C for over 24 hours in a controlled environment. After solvent evaporation, the substrates are separated to retrieve the resulting single-crystal $MAPbBr_3$ thin films. More details can be found in supplementary information.

**Device fabrication.** Firstly, 250-nm-thick $Si_3N_4$ is deposited on a 2-μm-thick copper sacrificial layer on the silicon substrate using plasma enhanced chemical vapor deposition (PECVD) method. After that, electron-beam lithography is performed on 200-nm-thick electron-beam resist ZEP 520A to define the photonic crystal patterns. These patterns are further transferred to the $Si_3N_4$ layer by dry etching. Next, the prepared photonic crystal is placed in a 99% $FeCl_3$ solution for wet etching, and after the copper sacrificial layer has been completely etched, the $Si_3N_4$ photonic crystals are released and then transferred to a PDMS stamp mounted on a high-precision piezo-stage for following measurements. More details can be found in supplementary information.

**Optical Characterization.** A home-built microscope is used to enable real-space and momentum-space resolved spectroscopy and imaging. The emission from the microcavity is collected using a 50× objective and then sent to a monochromator (Princeton Instruments Acton 2300i) with a liquid nitrogen–cooled charge-coupled device camera (Princeton Instruments, PIXIS-100B). The excitation source is a 500-nm non-resonant pulsed laser (pulse duration: 100 fs, repetition rate: 80 MHz). The repetition rate of the laser is further reduced to 1 MHz by a pulse picker to reduce the heat effect. Second-order quantum coherence is acquired using a HBT setup, where two fiber-coupled single-photon counting modules measure quantum correlations between arbitrary points in momentum space. More details can be found in supplementary information.

## DATA AVAILABILITY

All data supporting the figures in the main text or the supplementary materials are available upon request.

## CODE AVAILABILITY

The codes that support the findings of this study are available upon request

## METHODS-ONLY REFERENCES

No.

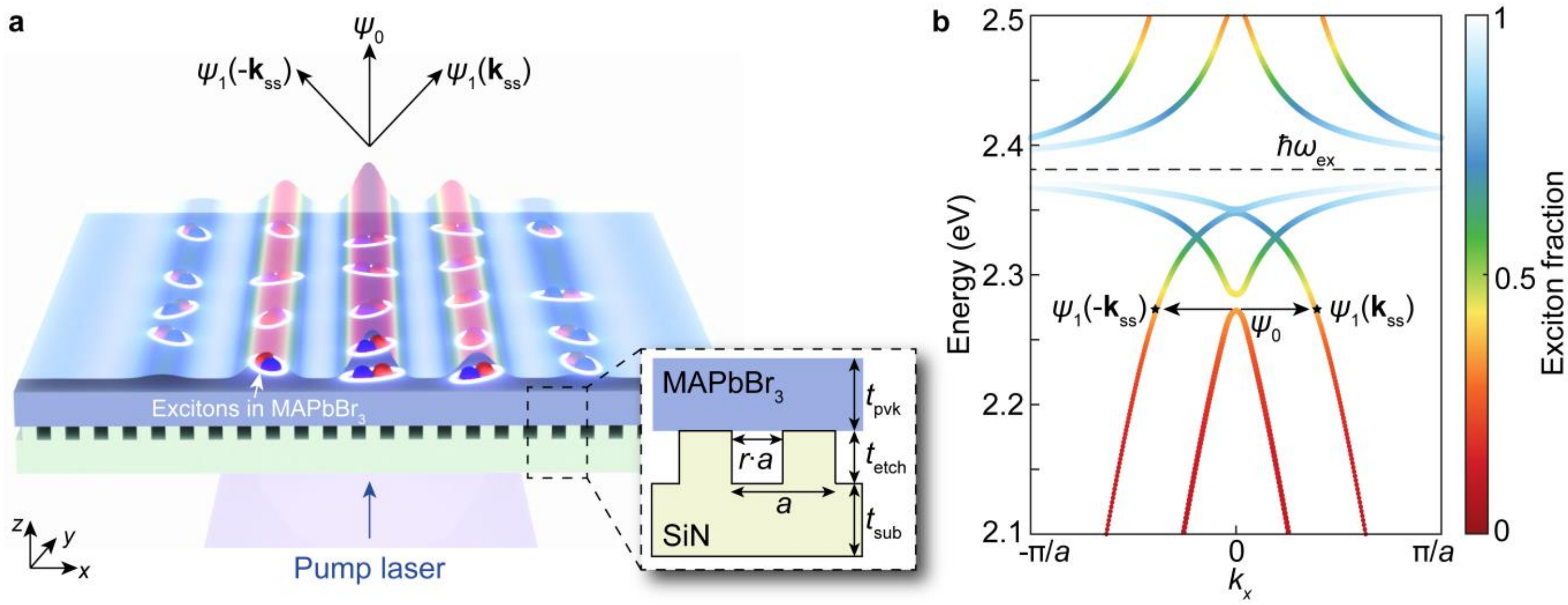

**Fig. 1 | Room-temperature supersolid in an exciton–polariton photonic-crystal device. a**, Schematic illustration of exciton–polariton supersolid at room temperature. The device integrates a room-temperature-stable perovskite ($MAPbBr_3$) into a dispersion-engineered $Si_3N_4$ photonic-crystal waveguide. Inset shows the cross-sectional device structure, with key device parameters: $MAPbBr_3$ thickness $t_{pvk}$ = 70 nm; $Si_3N_4$ etch depth $t_{etch}$ = 50 nm on a residual substrate thickness $t_{sub}$ = 200 nm; lattice constant $a$ = 290 nm; and duty cycle $r$ = 0.2. **b**, Theoretically calculated polariton dispersion. The photonic-crystal patterning folds the fundamental lower-polariton branch to the Brillouin zone center ($k_x = k_y = 0$), forming a high-$Q$, symmetry-protected BIC mode label as $\psi_0$. It also supports a symmetric pair of counter-propagating first-order modes $\psi_1(\pm\mathbf{k})$ at finite momenta about the Brillouin zone center. When $\psi_0$ is populated into condensate, the strong nonlinear interactions mediate a squeezing interaction between the $\psi_1(\pm\mathbf{k})$ modes, providing a parametric gain that eventually leads to a supersolid phase.

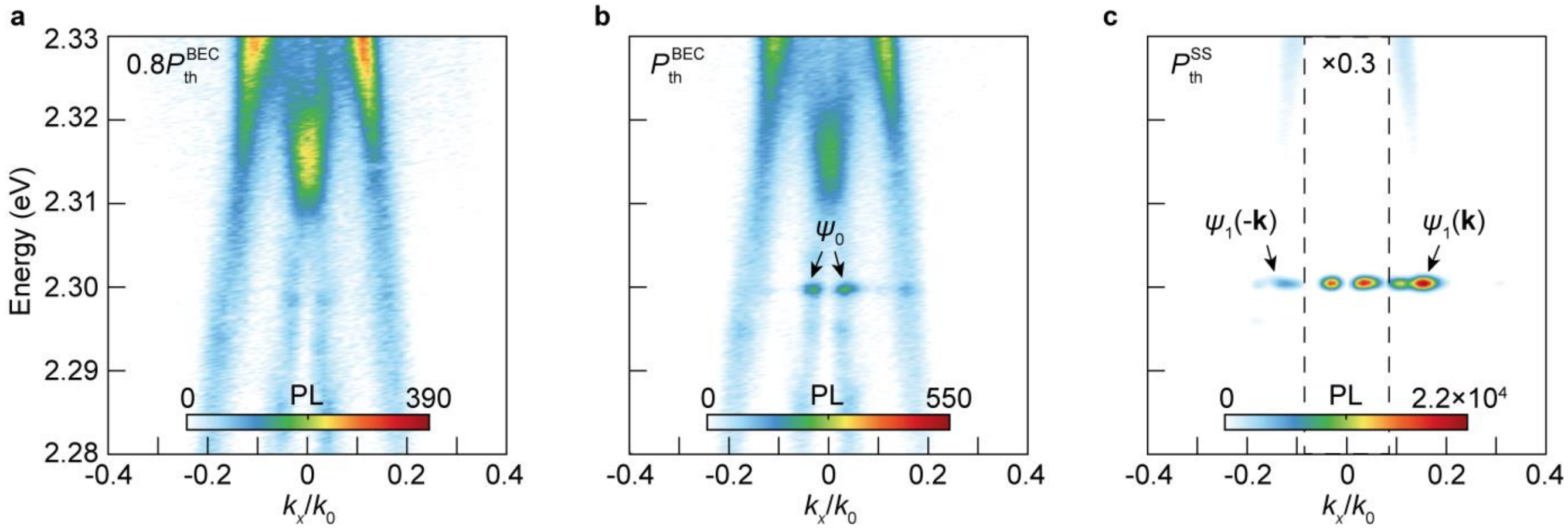

**Fig. 2 | Characterization of the supersolid phase in momentum space. a–c**, Angle-resolved photoluminescent spectra at pump intensity of $0.8P_{\mathrm{th}}^{\mathrm{BEC}}$ = 10.38 μJ/cm$^2$ (**a**), $P_{\mathrm{th}}^{\mathrm{BEC}}$ = 13.14 μJ/cm$^2$ (**b**), and $P_{\mathrm{th}}^{\mathrm{SS}}$ = 15.89 μJ/cm$^2$ (**c**). Below the condensation threshold, the device exhibits broad spontaneous emission from multiple polariton branches (**a**). The band-edge BIC mode possesses a negative effective mass, facilitating polariton accumulation as the pump intensity increases. At the BEC threshold $P_{\mathrm{th}}^{\mathrm{BEC}}$, a sharp condensate emerges in the high-$Q$ BIC mode $\psi_0$ (**b**). Upon further increasing the pump intensity to the supersolid threshold $P_{\mathrm{th}}^{\mathrm{SS}}$, distinct spectral peaks emerge at finite momenta $\pm\mathbf{k}_{\mathrm{ss}}$, signifying macroscopic occupation of the first-order modes $\psi_1(\pm\mathbf{k}_{\mathrm{ss}})$ (**c**). These peaks are iso-energetic with the condensate, in agreement with the theory that the parametric gain is maximized at zero energy detuning.

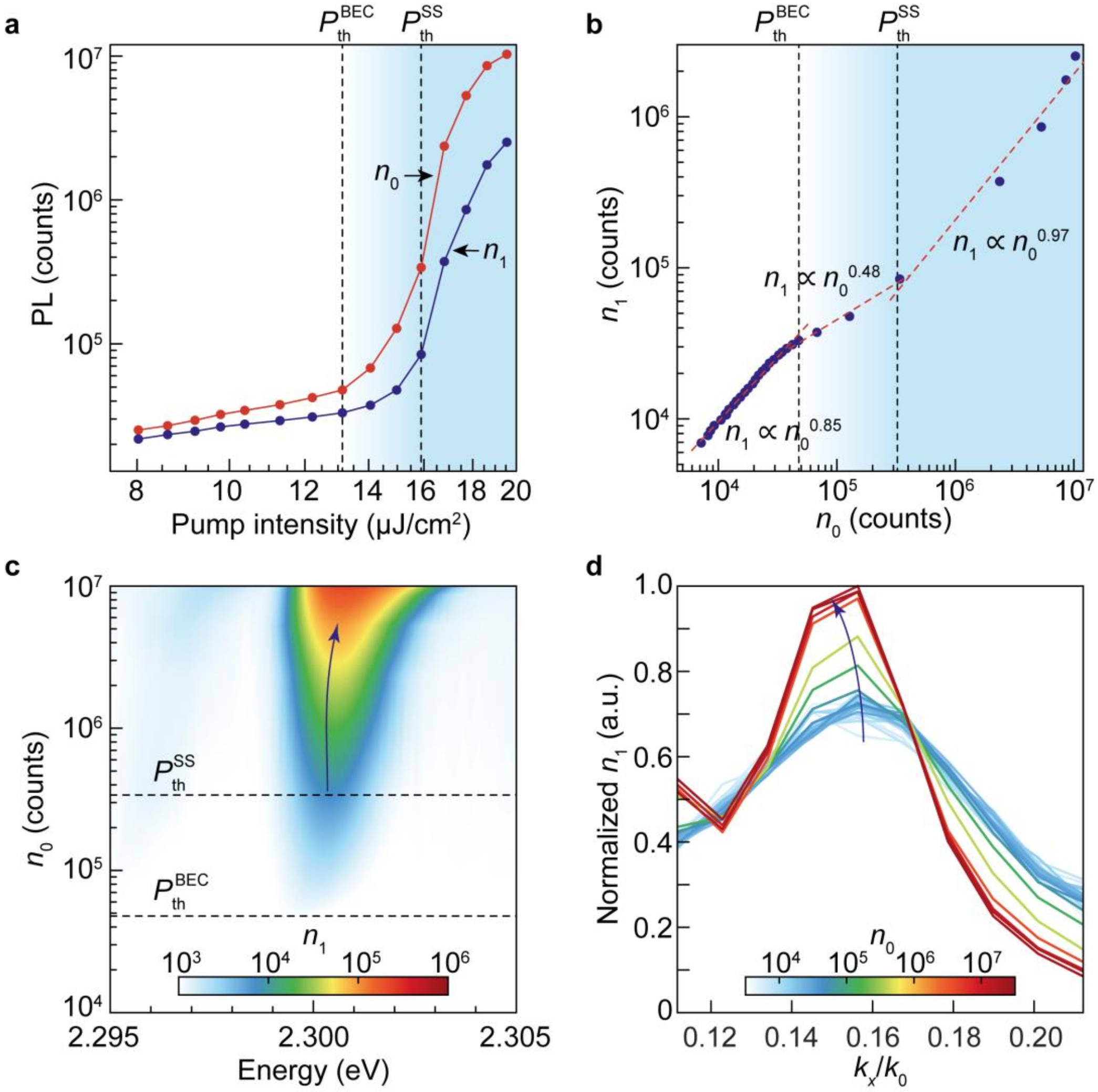

**Fig. 3 | Spontaneous emergence of non-rigid supersolid. a**, Measured emission intensity of the $\psi_0$ and $\psi_1(\pm\mathbf{k}_{ss})$ modes as a function of pump intensity. Condensation into $\psi_0$ mode occurs at $P_{th}^{BEC}$, followed by the phase transition to supersolid phase at $P_{th}^{SS}$. **b**, Supersolid component density $n_1$ as a function of the condensate density $n_0$. In the spontaneous emission regime, $n_1$ scales as $n_1 \propto n_0^{0.85}$. Between the two thresholds, $n_1$ exhibits sub-linear growth, scaling as $n_1 \propto n_0^{0.48}$. Above $P_{th}^{SS}$, the scaling becomes nearly linear, with $n_1 \propto n_0^{0.97}$, which is consistent with the theoretical prediction of a linear dependence $n_1 \propto \alpha_1 n_0$ when $\alpha_1 n_0 \gg \Gamma(\mathbf{k}_{ss})$. **c**, Energy spectra of $\psi_1(\pm\mathbf{k}_{ss})$ modes at different condensation density $n_0$. **d**, Momentum-space distribution of $\psi_1(\mathbf{k})$ for different condensation density $n_0$. The shifts in momentum space indicate the non-rigid nature of the supersolid phase.

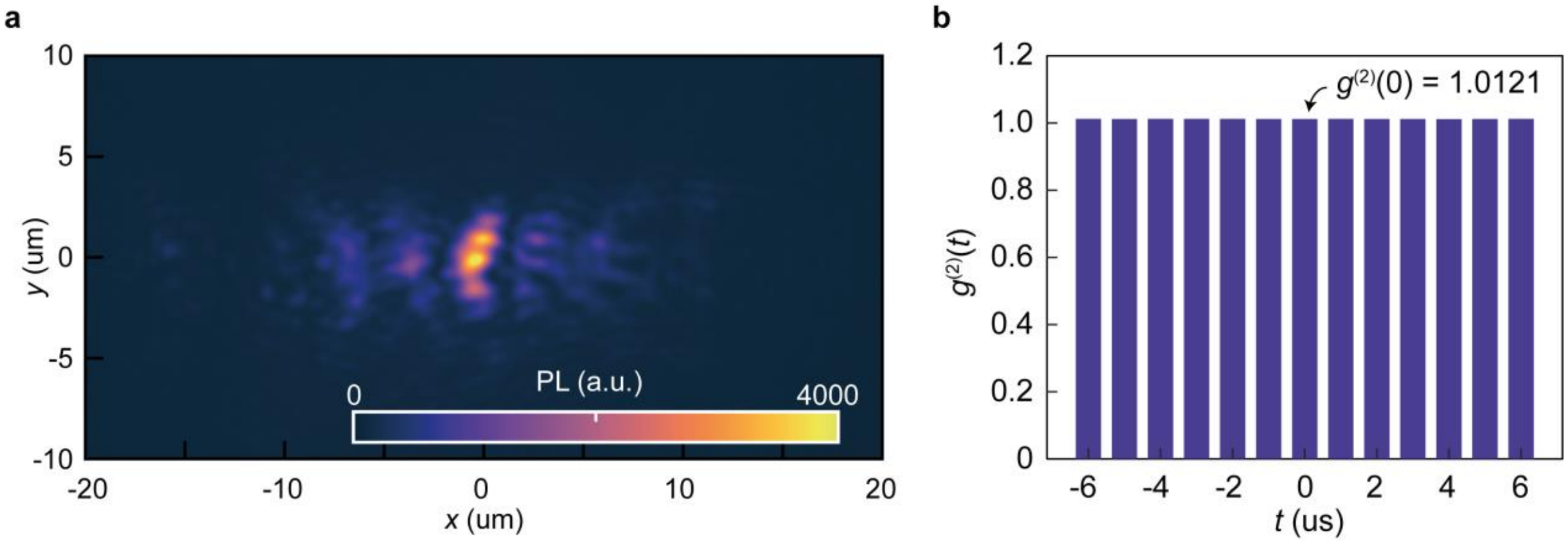

**Fig. 4 | Characterization of long-range orders of supersolids.** **a**, Real-space image of the supersolid arising from the coherent superposition of the macroscopically occupied $\psi_0$ and $\psi_1(\pm\mathbf{k}_{\mathrm{ss}})$ modes. **b**, Measured momentum-space second-order temporal coherence $g^2(\mathbf{k}_{\mathrm{ss}}, -\mathbf{k}_{\mathrm{ss}}; \tau)$ between the emission at $\pm\mathbf{k}_{\mathrm{ss}}$. At zero time delay, the cross-correlation $g^2(\mathbf{k}_{\mathrm{ss}}, -\mathbf{k}_{\mathrm{ss}}; 0) = 1.0121$ is very close to 1, providing clear evidence for quantum coherence between $\psi_1(\mathbf{k}_{\mathrm{ss}})$ and $\psi_1(-\mathbf{k}_{\mathrm{ss}})$ modes.